%% file: main.tex
\documentclass[12pt,a4paper]{article}
\pdfoutput=1
\usepackage{lineno}  
\usepackage{graphicx}  

\usepackage{graphpap}
\usepackage{rotating}

\usepackage{xspace} 
\usepackage{color}
\usepackage{colortbl}
\usepackage{amsmath} 
\usepackage{ifthen} 
\graphicspath{{./figs/}} 

\usepackage{multirow}

\usepackage{verbatim}

\newboolean{pdflatex}
\setboolean{pdflatex}{true} 
\newboolean{articletitles}
\setboolean{articletitles}{true} 

\newboolean{uprightparticles}
\setboolean{uprightparticles}{false} 

\usepackage{amssymb}
\usepackage{amsfonts}
\usepackage{upgreek} 

\usepackage{hyperref}    
\usepackage[all]{hypcap} 

\input{lhcb-symbols-def} 

\usepackage{cite} 
\usepackage{mciteplus}

\begin{document}

\renewcommand{\thefootnote}{\fnsymbol{footnote}}
\setcounter{footnote}{1}

\input{title-LHCb-PAPER}


\renewcommand{\thefootnote}{\arabic{footnote}}
\setcounter{footnote}{0}



\pagestyle{plain} 
\setcounter{page}{1}
\pagenumbering{arabic}


%

\input{introduction}

\input{detector}
\input{evsel}

\input{Nratio}

\input{effic}

\input{result}

\input{acknowledgements}

\addcontentsline{toc}{section}{References}
\bibliographystyle{LHCb}
\bibliography{main,local}

\end{document}

%% file: lhcb-symbols-def.tex
\def\lhcb {LHCb\xspace}
\def\ux85 {UX85\xspace}

\def\dzero  {D0\xspace}

\ifthenelse{\boolean{uprightparticles}}%
{

 \def\Pmu         {\ensuremath{\upmu}\xspace}

 \def\Ppi         {\ensuremath{\uppi}\xspace}

 \def\Ppsi        {\ensuremath{\uppsi}\xspace}

 \def\PDelta      {\ensuremath{\Delta}\xspace}                 
 \def\PXi      {\ensuremath{\Xi}\xspace}                 
 \def\PLambda      {\ensuremath{\Lambda}\xspace}                 
 \def\PSigma      {\ensuremath{\Sigma}\xspace}                 
 \def\POmega      {\ensuremath{\Omega}\xspace}                 
 \def\PUpsilon      {\ensuremath{\Upsilon}\xspace}                 
 
 \def\PB      {\ensuremath{\mathrm{B}}\xspace}                 
                  
 \def\PD      {\ensuremath{\mathrm{D}}\xspace}

 \def\PJ      {\ensuremath{\mathrm{J}}\xspace}                 
 \def\PK      {\ensuremath{\mathrm{K}}\xspace}

 \def\Pb      {\ensuremath{\mathrm{b}}\xspace}                 
 \def\Pc      {\ensuremath{\mathrm{c}}\xspace}                 
                  
 \def\Pe      {\ensuremath{\mathrm{e}}\xspace}

 \def\Pi      {\ensuremath{\mathrm{i}}\xspace}

 \def\Ps      {\ensuremath{\mathrm{s}}\xspace}

}
{

 \def\Pmu         {\ensuremath{\mu}\xspace}

 \def\Ppi         {\ensuremath{\pi}\xspace}

 \def\Ppsi        {\ensuremath{\psi}\xspace}                 
                  
 \mathchardef\PDelta="7101
 \mathchardef\PXi="7104
 \mathchardef\PLambda="7103
 \mathchardef\PSigma="7106
 \mathchardef\POmega="710A
 \mathchardef\PUpsilon="7107
                  
 \def\PB      {\ensuremath{B}\xspace}                 
                  
 \def\PD      {\ensuremath{D}\xspace}

 \def\PJ      {\ensuremath{J}\xspace}                 
 \def\PK      {\ensuremath{K}\xspace}

 \def\Pb      {\ensuremath{b}\xspace}                 
 \def\Pc      {\ensuremath{c}\xspace}                 
                  
 \def\Pe      {\ensuremath{e}\xspace}

 \def\Pi      {\ensuremath{i}\xspace}

 \def\Ps      {\ensuremath{s}\xspace}

}

\def\epem       {\ensuremath{\Pe^+\Pe^-}\xspace}

\def\mumu       {\ensuremath{\Pmu^+\Pmu^-}\xspace}

\def\squark    {\ensuremath{\Ps}\xspace}

\def\cquark    {\ensuremath{\Pc}\xspace}

\def\bquark    {\ensuremath{\Pb}\xspace}

\def\pion  {\ensuremath{\Ppi}\xspace}

\def\pim   {\ensuremath{\pion^-}\xspace}

\def\kaon  {\ensuremath{\PK}\xspace}
  \def\Kbar  {\kern 0.2em\overline{\kern -0.2em \PK}{}\xspace}

\def\Kz    {\ensuremath{\kaon^0}\xspace}
\def\Kzb   {\ensuremath{\Kbar^0}\xspace}
\def\KzKzb {\ensuremath{\Kz \kern -0.16em \Kzb}\xspace}
\def\Kp    {\ensuremath{\kaon^+}\xspace}
\def\Km    {\ensuremath{\kaon^-}\xspace}

\def\KpKm  {\ensuremath{\Kp \kern -0.16em \Km}\xspace}

\def\Kstarz  {\ensuremath{\kaon^{*0}}\xspace}

  \def\Dbar    {\kern 0.2em\overline{\kern -0.2em \PD}{}\xspace}
\def\D       {\ensuremath{\PD}\xspace}

\def\Dz      {\ensuremath{\D^0}\xspace}
\def\Dzb     {\ensuremath{\Dbar^0}\xspace}
\def\DzDzb   {\ensuremath{\Dz {\kern -0.16em \Dzb}}\xspace}
\def\Dp      {\ensuremath{\D^+}\xspace}
\def\Dm      {\ensuremath{\D^-}\xspace}

\def\DpDm    {\ensuremath{\Dp {\kern -0.16em \Dm}}\xspace}

\def\B       {\ensuremath{\PB}\xspace}
  \def\Bbar    {\kern 0.18em\overline{\kern -0.18em \PB}{}\xspace}

\def\Bu      {\ensuremath{\B^+}\xspace}

\def\Bp      {\ensuremath{\Bu}\xspace}

\def\Bd      {\ensuremath{\B^0}\xspace}
\def\Bs      {\ensuremath{\B^0_\squark}\xspace}

\def\jpsi     {\ensuremath{{\PJ\mskip -3mu/\mskip -2mu\Ppsi\mskip 2mu}}\xspace}

  \def\Y#1S{\ensuremath{\PUpsilon{(#1S)}}\xspace}

\def\BF         {{\ensuremath{\cal B}\xspace}}

\def\BR         {\BF}

\def\to                 {\ensuremath{\rightarrow}\xspace}

\def\CP                {\ensuremath{C\!P}\xspace}

\def\AT#1     {\ensuremath{A_{\mathrm{T}}^{#1}}\xspace}           

\def\C#1      {\ensuremath{\mathcal{C}_{#1}}\xspace}                       
\def\Cp#1     {\ensuremath{\mathcal{C}_{#1}^{'}}\xspace}                    
\def\Ceff#1   {\ensuremath{\mathcal{C}_{#1}^{\mathrm{(eff)}}}\xspace}        
\def\Cpeff#1  {\ensuremath{\mathcal{C}_{#1}^{'\mathrm{(eff)}}}\xspace}       
\def\Ope#1    {\ensuremath{\mathcal{O}_{#1}}\xspace}                       
\def\Opep#1   {\ensuremath{\mathcal{O}_{#1}^{'}}\xspace}                    

\newcommand{\tev}{\ensuremath{\mathrm{\,Te\kern -0.1em V}}\xspace}
\newcommand{\gev}{\ensuremath{\mathrm{\,Ge\kern -0.1em V}}\xspace}
\newcommand{\mev}{\ensuremath{\mathrm{\,Me\kern -0.1em V}}\xspace}
\newcommand{\kev}{\ensuremath{\mathrm{\,ke\kern -0.1em V}}\xspace}
\newcommand{\ev}{\ensuremath{\mathrm{\,e\kern -0.1em V}}\xspace}
\newcommand{\gevc}{\ensuremath{{\mathrm{\,Ge\kern -0.1em V\!/}c}}\xspace}
\newcommand{\mevc}{\ensuremath{{\mathrm{\,Me\kern -0.1em V\!/}c}}\xspace}
\newcommand{\gevcc}{\ensuremath{{\mathrm{\,Ge\kern -0.1em V\!/}c^2}}\xspace}
\newcommand{\gevgevcccc}{\ensuremath{{\mathrm{\,Ge\kern -0.1em V^2\!/}c^4}}\xspace}
\newcommand{\mevcc}{\ensuremath{{\mathrm{\,Me\kern -0.1em V\!/}c^2}}\xspace}

\def\mum  {\ensuremath{\,\upmu\rm m}\xspace}

\def\invfb   {\ensuremath{\mbox{\,fb}^{-1}}\xspace}

\newcommand{\chisq}{\ensuremath{\chi^2}\xspace}

\def\gsim{{~\raise.15em\hbox{$>$}\kern-.85em
          \lower.35em\hbox{$\sim$}~}\xspace}
\def\lsim{{~\raise.15em\hbox{$<$}\kern-.85em
          \lower.35em\hbox{$\sim$}~}\xspace}

\def\evtgen     {\mbox{\textsc{EvtGen}}\xspace}
\def\pythia     {\mbox{\textsc{Pythia}}\xspace}

\def\geant      {\mbox{\textsc{Geant4}}\xspace}

\def\tell1  {TELL1\xspace}
\def\ukl1   {UKL1\xspace}



%% file: title-LHCb-PAPER.tex

\begin{titlepage}
\pagenumbering{roman}

\vspace*{-1.5cm}
\centerline{\large EUROPEAN ORGANIZATION FOR NUCLEAR RESEARCH (CERN)}
\vspace*{1.5cm}
\hspace*{-0.5cm}
\begin{tabular*}{\linewidth}{lc@{\extracolsep{\fill}}r}
\ifthenelse{\boolean{pdflatex}}
{\vspace*{-2.7cm}\mbox{\!\!\!\includegraphics[width=.14\textwidth]{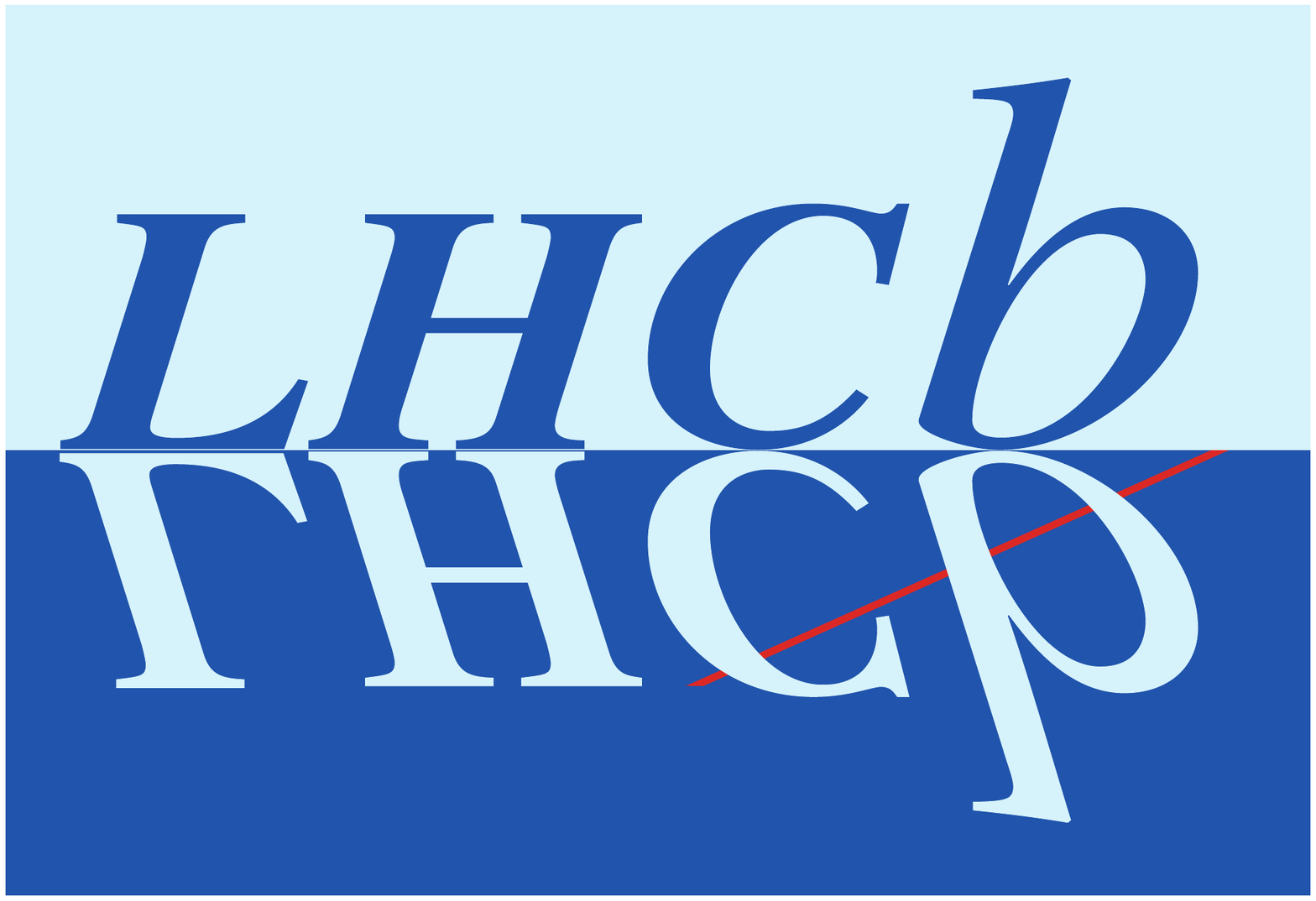}} & &}%
{\vspace*{-1.2cm}\mbox{\!\!\!\includegraphics[width=.12\textwidth]{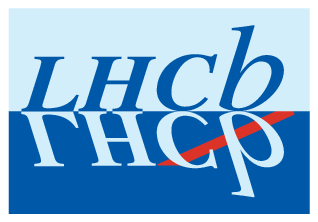}} & &}%
\\
 & & CERN-PH-EP-2012-113 \\  
 & & LHCb-PAPER-2012-010 \\  
 & & 4 May 2012 \\ 
 & & \\
\end{tabular*}

\vspace*{4.0cm}

{\bf\boldmath\huge
\begin{center}
    Measurement of relative branching fractions of \B decays to $\psi(2S)$ and $\jpsi$ mesons
\end{center}
}

\vspace*{2.0cm}

\begin{center}
The LHCb collaboration
\footnote{Authors are listed on the following pages.}
\end{center}

\vspace{\fill}

\begin{abstract}
  \noindent

The relative rates of \B-meson decays into $\jpsi$ and $\psi(2S)$ mesons are measured for the three decay modes in 
$pp$ collisions recorded with the LHCb detector. The ratios of branching fractions (\BR) are measured to be

\begin{equation*}
\begin{array}{lll}
\frac{\BR(\Bp\to \psi(2S) \Kp)}{\BR(\Bp\to\jpsi\Kp)} 	&=& 0.594 \pm 0.006\,(\mathrm{stat}) \pm 0.016\,(\mathrm{syst}) \pm 0.015\,(R_{\psi}), 
\\
\noalign{\vskip 3pt}
\frac{\BR(\Bd\to \psi(2S) \Kstarz)}{\BR(\Bd\to\jpsi\Kstarz)} 	&=& 0.476 \pm 0.014\,(\mathrm{stat}) \pm 0.010\,(\mathrm{syst}) \pm 0.012\,(R_{\psi}), 
\\
\noalign{\vskip 3pt}
\frac{\BR(\Bs\to \psi(2S)\phi)}{\BR(\Bs\to\jpsi\phi)} 	&=& 0.489 \pm 0.026\,(\mathrm{stat}) \pm 0.021\,(\mathrm{syst}) \pm 0.012\,(R_{\psi}),
\end{array}
\end{equation*}

\noindent where the third uncertainty is from the ratio of the $\psi(2S)$ and $\jpsi$ branching fractions to $\mumu$.

\end{abstract}

\vspace*{\fill}

\begin{center}
  Submitted to Eur. Phys. J. C
\end{center}

\vspace{\fill}

\end{titlepage}


\newpage
\setcounter{page}{2}
\mbox{~}
\newpage

\input{LHCb_authorlist.tex}

\cleardoublepage

%% file: LHCb_authorlist.tex
\centerline{\large\bf LHCb collaboration}
\begin{flushleft}
\small
R.~Aaij$^{38}$, 
C.~Abellan~Beteta$^{33,n}$, 
A.~Adametz$^{11}$, 
B.~Adeva$^{34}$, 
M.~Adinolfi$^{43}$, 
C.~Adrover$^{6}$, 
A.~Affolder$^{49}$, 
Z.~Ajaltouni$^{5}$, 
J.~Albrecht$^{35}$, 
F.~Alessio$^{35}$
% ...

{\footnotesize \it
% ...
$ ^{5}$Clermont Universit\'{e}, Universit\'{e} Blaise Pascal, CNRS/IN2P3, LPC, Clermont-Ferrand, France\\
$ ^{6}$CPPM, Aix-Marseille Universit\'{e}, CNRS/IN2P3, Marseille, France\\
% ...
$ ^{11}$Physikalisches Institut, Ruprecht-Karls-Universit\"{a}t Heidelberg, Heidelberg, Germany\\
% ...
$ ^{33}$Universitat de Barcelona, Barcelona, Spain\\
$ ^{34}$Universidad de Santiago de Compostela, Santiago de Compostela, Spain\\
$ ^{35}$European Organization for Nuclear Research (CERN), Geneva, Switzerland\\
% ...
$ ^{38}$Nikhef National Institute for Subatomic Physics, Amsterdam, The Netherlands\\
% ...
$ ^{43}$H.H. Wills Physics Laboratory, University of Bristol, Bristol, United Kingdom\\
% ...
$ ^{49}$Oliver Lodge Laboratory, University of Liverpool, Liverpool, United Kingdom\\
% ...
$ ^{n}$LIFAELS, La Salle, Universitat Ramon Llull, Barcelona, Spain\\
% ...
}
\end{flushleft}

%% file: introduction.tex
%

\section{Introduction}
\label{sec:Introduction}

Decays of \B mesons to two-body final states containing a charmonium resonance such 
as a $\jpsi$ or $\psi(2S)$ offer a powerful way of studying  electroweak transitions. 
Such decays probe charmonium properties and play a role in the study of \CP~violation 
and mixing in the neutral \B~system~\cite{Bigi:1981qs}.

The relative branching fractions of \Bp, \Bd and \Bs mesons into 
$\jpsi$ and $\psi(2S)$ mesons have previously been studied by both the CDF and \dzero 
collaborations~\cite{CDF1,CDF2,D0}. 
Since the current experimental results for the study of \CP violation 
in $\Bs$ mixing using the $\Bs\to\jpsi\phi$~decay~\cite{BsCDF,Abazov:2011ry,LHCb-PAPER-2011-021} 
are statistically limited, it is important to establish other channels 
where this analysis can be done. One such channel is the $\Bs\to\psi(2S)\phi$ 
decay. 

In this paper, measurements of the ratios of the branching fractions of \B mesons decaying 
to $\psi(2S)X$ and $\jpsi X$ are reported,
where \B denotes a \Bp, \Bd or \Bs meson (charge conjugate decays are implicitly included) 
and $X$ denotes a $\Kp$, $\Kstarz$ or $\phi$ meson. 
The data were collected by the \lhcb experiment in~$pp$~collisions 
at the centre-of-mass energy $\sqrt{s} = 7\tev$ during 
2011 and correspond to an integrated luminosity of 0.37\invfb.

%% file: detector.tex
\section{Detector description}
\label{sec:Detector}

The \lhcb detector~\cite{Alves:2008zz} is a single-arm forward
spectrometer covering the pseudorapidity range $2<\eta <5$, designed
for the study of \bquark- and \cquark-hadrons. The
detector includes a high precision tracking system consisting of a
silicon-strip vertex detector surrounding the $pp$ interaction region,
a large-area silicon-strip detector located upstream of a dipole
magnet with a bending power of about $4{\rm\,Tm}$, and three stations
of silicon-strip detectors and straw drift-tubes placed
downstream. The combined tracking system has a momentum resolution
$\Delta p/p$ that varies from 0.4\% at 5\gevc to 0.6\% at 100\gevc,
and an impact parameter resolution of 20\mum for tracks with high
transverse momentum. 
Data were taken with both magnet polarities to reduce systematic effects due to detector asymmetries.
Charged hadrons are identified using two
ring-imaging Cherenkov (RICH) detectors. 
Photon, electron and hadron
candidates are identified by a calorimeter system consisting of
scintillating-pad and pre-shower detectors, and electromagnetic
and hadronic calorimeters. Muons are identified by a muon
system composed of alternating layers of iron and multiwire
proportional chambers. The trigger consists of a hardware 
stage based on information from the calorimeter and muon systems, 
followed by a software stage which applies a full event 
reconstruction.

Events with a~$\jpsi\to\mumu$~final state are triggered using 
two hardware trigger
decisions: the single-muon decision, which requires one muon
candidate with a transverse momentum $p_{\mathrm{T}}$~larger 
than 1.5~$\mathrm{GeV}/c$,
and the di-muon decision, which requires two muon candidates
with transverse momenta 
$p_{\mathrm{T}_1}$ and 
$p_{\mathrm{T}_2}$ 
satisfying the relation
$\sqrt{p_{\mathrm{T}_1} \cdot p_{\mathrm{T}_2}}> 1.3~\mathrm{GeV}/c$.
The di-muon trigger decision in the software trigger
requires muon pairs of opposite charge with $p_{\mathrm{T}}>500~\mathrm{MeV}/c$, 
forming a common vertex and with an invariant mass in excess of $2.9~\mathrm{GeV}/c^2$. 

%% file: evsel.tex
%

\newcommand{\tis}     {\ensuremath{\mathrm{TIS}}}
\newcommand{\nottis}  {\ensuremath{\overline{\mathrm{TIS}}}}
\newcommand{\tos}     {\ensuremath{\mathrm{TOS}}}
\newcommand{\nottos}  {\ensuremath{\overline{\mathrm{TOS}}}}

\section{Event selection}
\label{sec:EventSelection}

In this analysis, the decays $\Bp\to\psi\Kp$($\Bd\to\psi\Kstarz$, $\Bs\to\psi\phi$) are 
reconstructed, where $\psi$ represents $\psi(2S)$ or 
$\jpsi$, reconstructed in the $\psi\to\mumu$ decay modes. A $\Kp$($\Kstarz$, $\phi$) candidate 
is added to the di-muon pair to form a $\Bp$($\Bd$, $\Bs$ ) candidate. 

The starting point of the analysis is the reconstruction of either a $\jpsi$ or $\psi(2S)$ meson 
decaying into a di-muon pair. 
Candidates are formed from pairs of opposite sign tracks that both have a transverse momentum 
larger than 500\mevc. Good reconstruction quality is 
assured by requiring the $\chisq$ per degree of freedom of the track fit to satisfy $\chisq/\rm{ndf}<5$. 
Both tracks must be identified as muons. This is achieved by requiring the muon identification variable, 
the difference in logarithm of the likelihood of the muon and hadron hypotheses~\cite{LHCb-PROC-2011-008} 
provided by the muon detection system, to 
satisfy $\Delta\log\mathcal{L}^{\mu-h} > -5$. 
The muons are required to form a common vertex of good quality ($\mathrm{\chisq_{vtx}}<20$). 
The resulting di-muon candidate is required to have decay length significance from its 
associated primary vertex greater than 5 and 
have an invariant mass between 3020 and 3135\mevcc in the case of a $\jpsi$ candidate or 
between 3597 and 3730\mevcc for a $\psi(2S)$ 
candidate. These correspond to [$-5\sigma$; $3\sigma$] windows around the nominal mass. 
The asymmetric window allows for the QED radiative tail.

The selected $\jpsi$ and $\psi(2S)$ candidates are then combined with a $\Kp$, $\Kstarz$ 
or $\phi$ to create \B meson 
candidates. Only the $\Kstarz\to\Kp\pim$ and $\phi\to\Kp\Km$ decay modes are considered. 
Pion-kaon separation is provided by the ring-imaging Cherenkov detectors. To identify kaons 
the difference in logarithm of the 
likelihood of the kaon and pion hypotheses~\cite{LHCb-PROC-2011-008} is required to 
satisfy $\Delta\log\mathcal{L}^{\kaon-\pion} > -5$.
In the case of pions the difference in logarithm of the likelihood of the pion and kaon 
hypotheses~\cite{LHCb-PROC-2011-008} is required to satisfy 
$\Delta\log\mathcal{L}^{\pion-\kaon} > -5$.
As in the case of muons, a cut is applied on the track $\mathrm{\chi^2/ndf}$ provided by 
the track fit at 5. The kaons and pions are 
required to have a transverse momentum larger than 250\mevc and to have an impact parameter 
significance with respect to any primary vertex larger 
than 2. 
In the $\Bd$ channel, the mass of the kaon and pion system is required 
to be $842<M_{\Kp\pim}<942\mevcc$ and in the \Bs channel the mass of the 
kaon pair is required to be $1010<M_{\Kp\Km}<1030\mevcc$.

In addition, we require the decay time of the \B candidate ($c\tau$) to be larger 
than 100\mum to reduce the large combinatorial background from 
particles produced in the primary $pp$ interaction. 
A global refit of the three-prong (four-prong) combination is performed with a 
primary vertex constraint 
and with the di-muon pair mass constrained to the nominal value~\cite{Nakamura:2010zzi} using 
the Decay Tree Fit (DTF) procedure~\cite{DTF}. 
The reduced $\chi^2$ of this fit ($\mathrm{\chi^2_{DTF}/ndf}$) is required 
to be less than 5, where the DTF algorithm takes into account the number of decay 
products to determine the number of degrees of freedom. 
The \Bp candidates, where a muon from the $\psi(2S)\to\mumu$ decay is reconstructed 
as both muon and kaon, are 
removed by requiring the angle between the same sign muon and kaon to be greater than 3 mrad.

\begin{figure}[t]
  \setlength{\unitlength}{1mm}
  \centering
  \begin{picture}(160,110)
    \put(0,80){
      \includegraphics*[width=160mm,
      ]{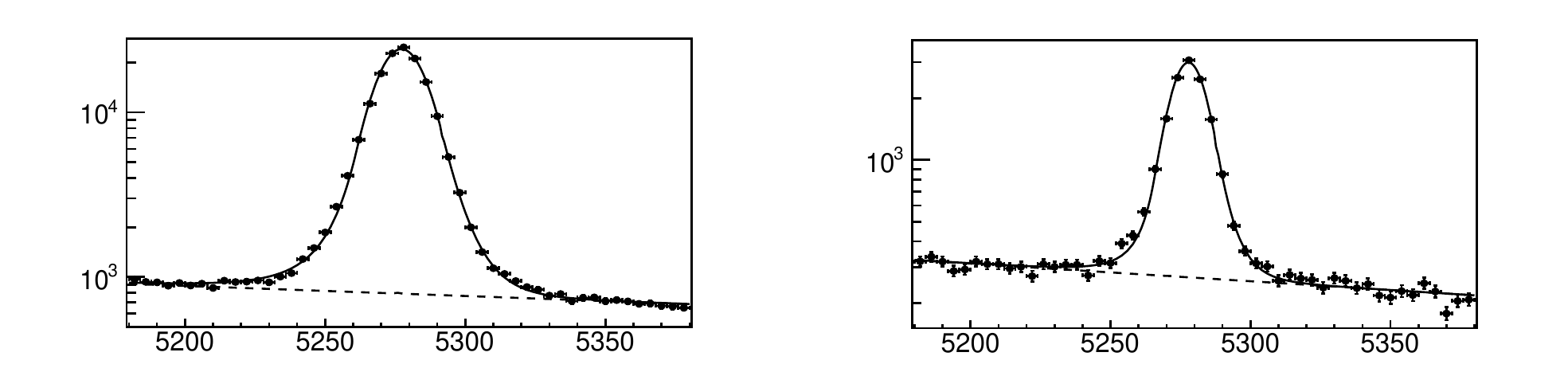}
    }
    \put(0,39){
      \includegraphics*[width=160mm,
      ]{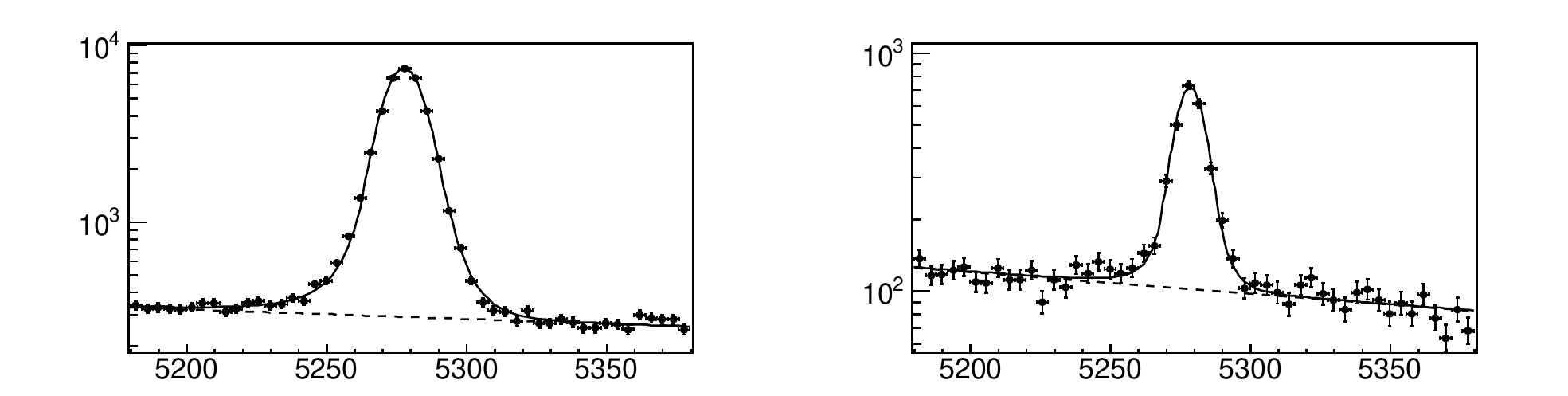}
    }
    \put(0,0){
      \includegraphics*[width=160mm,
      ]{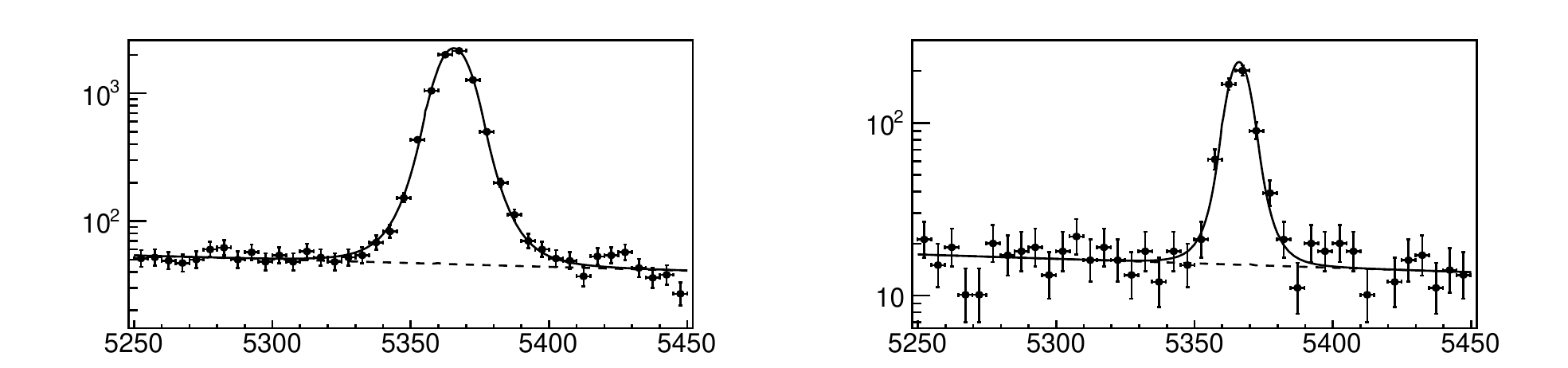}
    }

    \put(20,  107) 	{ \small        \text{LHCb} }
    \put(100,  107)	{ \small	\text{LHCb} }
    \put(20,  67) 	{ \small	\text{LHCb} }
    \put(100,  67) 	{ \small        \text{LHCb} }
    \put(20,  27) 	{ \small        \text{LHCb} } 
    \put(100,  27) 	{ \small        \text{LHCb} }

    \put(65,105){(a)}
    \put(145,105){(b)}
    \put(65,65){(c)}
    \put(145,65){(d)}
    \put(65,25){(e)}
    \put(145,25){(f)}

    \put(40,-2)	{\footnotesize $\mathrm{M}_{\jpsi\Kp\Km}	~[\mevcc]$}
    \put(118,-2){\footnotesize $\mathrm{M}_{\psi(2S)\Kp\Km}	~[\mevcc]$}
    \put(41,37)	{\footnotesize $\mathrm{M}_{\jpsi\Kp\pim}	~[\mevcc]$}
    \put(119,37){\footnotesize $\mathrm{M}_{\psi(2S) \Kp\pim}	~[\mevcc]$}
    \put(45,78)	{\footnotesize $\mathrm{M}_{\jpsi\Kp}		~[\mevcc]$}
    \put(122,78){\footnotesize $\mathrm{M}_{\psi(2S)\Kp}	~[\mevcc]$}

    \begin{sideways}
    \put(80,90){ \footnotesize Candidates/(4\,\mevcc)}
    \put(80,10)	{ \footnotesize Candidates/(4\,\mevcc)}
    \put(0,90)	{ \footnotesize Candidates/(5\,\mevcc)}
    \put(0,10)	{ \footnotesize Candidates/(5\,\mevcc)}
    \put(40,90){ \footnotesize Candidates/(4\,\mevcc)}
    \put(40,10)	{ \footnotesize Candidates/(4\,\mevcc)}
    \put(45,97)	{ \phantom}
    \end{sideways}

  \end{picture}
  \caption {\small Mass distributions of (a) $\Bp\to\jpsi\Kp$, 
        (b) $\Bp\to\psi(2S)\Kp$, (c) $\Bd\to\jpsi\Kp\pim$, (d) $\Bd\to\psi(2S)\Kp\pim$, 
        (e) $\Bs\to\jpsi\Kp\Km$ and (f) $\Bs\to\psi(2S)\Kp\Km$. The total fitted function (solid)
         and the combinatorial background (dashed) are shown. 
         The variation in resolution of the different modes is fully consistent 
         with the energy released in the decays and in agreement with simulation.}
  \label{fig:MassDist}
\end{figure}

%% file: Nratio.tex
%

\section{ Measurement of $\boldsymbol{ N_{\psi(2S)X} / N_{\jpsi X}}$ }
\label{sec:Nratio}
The mass distributions for selected candidates are shown in Fig.~\ref{fig:MassDist}. 
The number of the $\Bp\to\psi\Kp$ candidates is estimated by 
performing an unbinned maximum likelihood fit. 
The same procedure is used to determine the number of the $\Bd\to\psi\Kp\pim$ candidates 
in a $842~<~M_{\Kp\pim}~<~942\mevcc$ mass window and the number of the $\Bs\to\psi\Kp\Km$ candidates 
in a $1010~<~M_{\Kp\Km}~<~1030\mevcc$ mass window.
The number of signal candidates is determined by fitting a double-sided
Crystal Ball function~\cite{Skwarnicki:1986xj,LHCb-PAPER-2011-013} for signal together 
with an exponential function to model the background. The tail parameters
of the Crystal Ball function are fixed to values determined from simulation.

The \Bd mass distributions include the contributions from resonant decays ($\Bd\to\psi\Kstarz$), 
non-resonant 
decays ($\Bd\to\psi\Kp\pim$) and combinatorial background. 
The contributions from resonant and non-resonant modes are separated with the {\em sPlot} 
technique~\cite{sPlot}.
The $\Kp\pim$ invariant mass is used as a discriminating variable to unfold the \Bd mass 
distribution of non-$\Kstarz$ 
$\Kp\pim$ combinations. A fit is then performed to the unfolded \Bd distribution, 
which contains both non-resonant $\Bd\to\psi\Kp\pim$ decays and background, 
to determine the number of non-resonant decays. The final number of resonant decays is calculated 
by subtracting the number of non-resonant decays from the total number of decays. 
For the \Bs modes the number of non-resonant decays are obtained by a similar procedure using 
the $\Kp\Km$ invariant mass as the discriminating variable. 
The signal yields and their ratios are summarized in Table~\ref{table:Nevents}.

\begin{table}[tb]
\caption{\small Summary of the signal yields for the six \B modes considered and the ratios 
               of the number of $\jpsi$ and $\psi(2S)$ decays: 
               $N^{\rm{total}}$ is the summed signal yield for resonant and non-resonant modes, 
               $N^{\rm{non-res}}$ is the signal yield for non-resonant modes only and 
               $N^{\rm{res}}_{\psi X}$ is the signal yield for resonant decays 
               (through \Kstarz or $\phi$). The uncertainties are statistical only. }
\begin{center}
\begin{tabular}{l|c|c|c|c}
	\B decay modes 
    & $N^{\rm{total}}$ 
    & $N^{\rm{non-res}}$ 
    & $N^{\rm{res}}$ 
    & $N^{\rm{res}}_{\psi(2S)X}/N^{\rm{res}}_{\jpsi X}$ \\
	\hline
	$\Bp\to\jpsi\Kp$ 	
    & $141,769\pm410$ 		
    & {---} 
    & $141,769\pm410$ 	
    & \multirow{2}{*}{$0.0857\pm0.0009$} \\
	$\Bp\to\psi(2S)\Kp$ 	
    & \;\:$12,154\pm130$ 
    & {---}
    & \;\:$12,154\pm130$ 	& \\
	\hline
	$\Bd\to\jpsi\Kp\pim$ 	
    & \;\:$35,770\pm207$ 
    & $1,253\pm30$
    & \;\:$34,517\pm209$
    & \multirow{2}{*}{$0.0612\pm0.0018$} \\
	$\Bd\to\psi(2S)\Kp\pim$ 
    & \;\:$2,223\pm60$	
    & \quad$112\pm12$& \;\:$2,111\pm61$	& \\
	\hline
	$\Bs\to\jpsi\Kp\Km$ 	
    & \;\:$7,654\pm92$	
    & \quad\;\:$66\pm13$& \;\:$7,588\pm93$	
    & \multirow{2}{*}{$0.0652\pm0.0034$} 
\\
	$\Bs\to\psi(2S)\Kp\Km$	
    & \quad\quad$495\pm25$\;\:	
    & \quad\;\:$0^{+1}_{-0}$	
    & \quad\quad$495\pm25$\;\:	& \\
\end{tabular}
\normalsize
\end{center}
\label{table:Nevents}
\end{table}

%% file: effic.tex
%

\section{Efficiencies and systematic uncertainties}
\label{sec:Efficiency}

The branching fraction ratio is calculated using 

\begin{equation}
\frac{\BR(B \to \psi(2S) X)}{\BR (B\to\jpsi X)} = 
\frac{N^{\rm{res}}_{\psi(2S) X}}{N^{\rm{res}}_{\jpsi X}} \times
\frac{\varepsilon_{\jpsi X}}{\varepsilon_{\psi(2S) X}}
\times \frac{\BR(\jpsi\to\mumu)}{\BR(\psi(2S)\to\mumu)} ,
\label{eq:overall}
\end{equation}

\noindent where $N^{\mathrm{res}}$ is the number of signal 
candidates and $\varepsilon$ is the overall efficiency.

The overall efficiency is the product of the geometrical acceptance of the detector, 
the combined reconstruction and selection 
efficiency, and the trigger efficiency. The efficiency ratio is estimated 
using simulation for all six modes. 
The simulation samples used are based on 
the \pythia~6.4 generator~\cite{Sjostrand:2006za} configured with 
the parameters detailed in Ref.~\cite{LHCb-PROC-2011-005}. 
Final state QED radiative corrections are included using 
the {\sc{Photos}} package~\cite{photos}.
The \evtgen~\cite{Lange:2001uf} and  \geant~\cite{Agostinelli:2002hh} packages 
are used to generate hadron decays and simulate interactions in the detector, 
respectively.
The digitized output is passed through a full simulation of both the 
hardware and software trigger and then reconstructed in the same way as the data.

The overall efficiency ratio is $0.901\pm0.016$, $1.011\pm0.014$ and 
$0.994\pm0.014$ for the \Bp, the \Bd and the \Bs channels respectively. 
Since the selection criteria for $B\to\jpsi X$ and $B\to\psi(2S) X$ decays are identical, 
the ratio of efficiencies is expected to be close to unity. 
The deviation of the overall efficiency ratio from unity in the case of the 
$\Bp\to\psi\Kp$ decays is due to the difference between the $p_{\rm{T}}$ spectra 
of muons for the $\jpsi$ and $\psi(2S)$ 
decays. For the \Bd and \Bs channels this difference is small. 
It has been checked that the behaviour of the efficiencies of 
all selection criteria is consistent in the data and simulation.

Since the decay products in each of the pairs of channels considered 
have similar kinematics, most uncertainties cancel in the ratio. 
The different contributions to the systematic uncertainties affecting 
this analysis are discussed in the following and summarized in 
Table~\ref{table:SysFitUnc}.

\begin{table}[tb]
\caption{\small Systematic uncertainties (in \%) on the relative branching fractions.}
\begin{center}
\begin{tabular}{l|c|c|c}

        Source          & \Bp channel & \Bd channel & \Bs channel \\
        \hline

        non-resonant decays		& ---   & $1.5$  		& $3.4$  \\%
        data-simulation agreement	& $1.7$ & $0.5$ 		& $2.0$ \\ %

        magnet polarity			& $1.4$ & $0.6$ 		& $0.7$ \\ %
        finite simulation sample size 	& $0.3$ & $0.5$ 		& $0.6$ \\ %
        trigger				& $1.1$ & $1.1$ 		& $1.1$ \\ %

        background shape		& $0.6$ & $0.2$ 		& $0.2$  \\ %
        signal shape 			& $0.7$ & $0.8$ 		& $0.5$  \\ %

        angular distribution		& ---   & $\!\!\!\!\!\!<0.1$ 	& $0.6$  		\\ %
        particle misidentification  	& $0.4$ & $\!\!\!\!\!\!<0.1$ 	& $\!\!\!\!\!\!<0.1$ 	\\ %

        \hline
        Sum in quadrature		& $2.7$ & $2.2$ & $4.3$ \\

\end{tabular}
\end{center}
\label{table:SysFitUnc}
\end{table}

The dominant source of systematic uncertainty arises from the subtraction of 
the non-resonant components in the $\Bd$ and the $\Bs$
decays. The non-resonant background is studied with two alternative methods. 
First, determining the number 
of $\Bd_{(s)}\to\psi\Kstarz(\phi)$ decays directly using the {\em sPlot} technique 
by unfolding 
and fitting the $\Bd_{(s)}$ mass distribution of candidates containing 
genuine $\Kstarz(\phi)$ resonances. 
Second, using  the $\Bd_{(s)}$ mass distribution as the discriminating 
variable to unfold the  
$\Kp\pim (\Kp\Km)$ mass distribution of genuine $\Bd_{(s)}$ candidates and fitting  
this distribution to determine the number of non-resonant decays. 
The corresponding uncertainties are 
found to be 1.5\% in the \Bd channel and 3.4\% in the \Bs channel.

The other important source of uncertainty arises from the estimation of 
the efficiencies due to the potential disagreement between data and 
simulation. This is studied by varying the selection criteria in data and simulation.
The corresponding uncertainties are found to be 1.7\% in the \Bp channel, 
0.5\% in the \Bd channel and 2.0\% in the \Bs channel.
The observed difference in the efficiency ratios for the two magnet polarities 
is conservatively taken as an estimate of the systematic 
uncertainty. This is 1.4\% in the \Bp channel, 0.6\% in the \Bd channel 
and 0.7\% in the \Bs channel.

The trigger is highly efficient in selecting \B meson decays with two muons in the final state.
For this analysis the di-muon pair is required to trigger the event. 
Differences in the trigger efficiency between data and simulation are studied 
in the data using events which were triggered independently on 
the di-muon pair \cite{LHCb-PUB-2011-016}. Based on these studies, an uncertainty of 1.1\% is assigned.

A further uncertainty arises from the imperfect knowledge of the shape of 
the signal and background in the \B meson mass distribution. To estimate this 
effect, a linear and a quadratic function are considered as alternative models 
for the background mass distribution.
In addition, a double Gaussian shape and a sum of double-sided Crystal Ball 
and Gaussian shapes are used as
alternative models for the signal shape.
The maximum observed change in the ratio of yields in the $\psi(2S)$ 
and $\jpsi$ modes is taken as systematic uncertainty.

The central value of the relative efficiency is determined by assuming 
that the angular distribution of the $B\to\psi(2S)X$ decay
is the same as that of the $B\to\jpsi X$. 
The systematic uncertainty due to the unknown polarization of 
the $\psi(2S)$ in the \B meson decays is estimated as follows. 
The simulation samples were re-weighted to match the angular distributions 
found from the data and the relative efficiency was recalculated. 
The difference between the baseline analysis and the re-weighted simulation is taken 
as the systematic uncertainty, as shown in Table~\ref{table:SysFitUnc}.

Finally, the uncertainty due to potential contribution from 
the Cabibbo-suppressed mode with a \pion misidentified as \kaon is found to be 0.4\% in 
the \Bp channel and negligible in the \Bd and \Bs channels. The uncertainty due 
to the cross-feed between \Bd and \Bs channels with a \pion misidentified as \kaon 
(or a \kaon misidentified as \pion) is negligible.

%% file: result.tex
%

\section{Results}
\label{sec:Result}

Since the di-electron branching fractions are measured more precisely than 
those of the di-muon decay modes, we assume lepton universality and take 
$R_{\psi}=\BR(\jpsi\to\mumu)/\BR(\psi(2S)\to\mumu) = \BR(\jpsi\to\epem)/\BR(\psi(2S)\to\epem) = 7.69\pm0.19$~\cite{Nakamura:2010zzi}. 
The results are combined using Eq.~\ref{eq:overall} to give 

$$
\begin{array}{lll} 
\frac{ \BR(\Bp\to\psi(2S)\Kp)}{ \BR(\Bp\to\jpsi\Kp)}	
&=& 0.594~\pm0.006\,({\rm stat})\pm0.016\,({\rm syst})\pm0.015\,(R_{\psi}), \\
\noalign{\vskip 3pt}
\frac{ \BR(\Bd\to\psi(2S)\Kstarz)}{ \BR(\Bd\to\jpsi\Kstarz)}	
&=& 0.476~\pm0.014\,({\rm stat})\pm0.010\,({\rm syst})\pm0.012\,(R_{\psi}), \\
\noalign{\vskip 3pt}
\frac{ \BR(\Bs\to\psi(2S)\phi)}{ \BR(\Bs\to\jpsi\phi)}	
&=& 0.489~\pm0.026\,({\rm stat})\pm0.021\,({\rm syst})\pm0.012\,(R_{\psi}), \\
\end{array}
$$
\noindent where the first uncertainty is statistical, the second is 
systematic and the third is the uncertainty on the 
$R_{\psi}$~value~\cite{Nakamura:2010zzi}.

The resulting branching fraction ratios are compatible with, 
but significantly more precise than, the current world averages of 
$\BR(\Bp\to\psi(2S)\Kp)/\BR(\Bp\to\jpsi\Kp)=0.60\pm0.07$ and 
$\BR(\Bs\to\psi(2S)\phi)/\BR(\Bs\to\jpsi\phi)=0.53\pm0.10$~\cite{Nakamura:2010zzi} and the CDF result 
of $\BR(\Bd\to\psi(2S)\Kstarz)/\BR(\Bd\to\jpsi\Kstarz)=0.515\pm0.113\pm0.052$~\cite{CDF1}. 
The $\Bs\to\psi(2S)\phi$ decay is particulary interesting 
since, with more data, it can be used for the measurement of \CP violation in \Bs mixing.

%% file: acknowledgements.tex
\section*{Acknowledgements}

\noindent We express our gratitude to our colleagues in the CERN accelerator
departments for the excellent performance of the LHC. We thank the
technical and administrative staff at CERN and at the LHCb institutes,
and acknowledge support from the National Agencies: CAPES, CNPq,
FAPERJ and FINEP (Brazil); CERN; NSFC (China); CNRS/IN2P3 (France);
BMBF, DFG, HGF and MPG (Germany); SFI (Ireland); INFN (Italy); FOM and
NWO (The Netherlands); SCSR (Poland); ANCS (Romania); MinES of Russia and
Rosatom (Russia); MICINN, XuntaGal and GENCAT (Spain); SNSF and SER
(Switzerland); NAS Ukraine (Ukraine); STFC (United Kingdom); NSF
(USA). We also acknowledge the support received from the ERC under FP7
and the Region Auvergne.